\documentclass[11pt]{article}
\usepackage{moriond, epsfig, fancyheadings}
\usepackage{graphicx}
\usepackage{natbib}
\def\be{\begin{equation}}
\def\ee{\end{equation}}
\def\bea{\begin{eqnarray}}
\def\eea{\end{eqnarray}}

\begin{document}
\vspace*{1cm}

\thispagestyle{fancy}
\lhead{Accepted for publication in Faraday Discussions 133 (FD133), Royal Society of Chemistry, 2006, 
DOI:10.1039/b601805j}

\title{Extragalactic chemistry of molecular gas: lessons from the local universe}
\author{S.~Garc\'{i}a-Burillo$^{1}$, A.~Fuente$^{1}$, J.~Mart\'{\i}n-Pintado$^{2}$, A.~Usero$^{1}$, J.~Graci\'a-Carpio$^{1}$, P.~Planesas$^{1}$}

\address{$^{1}$Observatorio Astron\'omico Nacional-OAN, Observatorio de Madrid, Alfonso XII, 3, E-28014, Madrid, SPAIN}
\address{$^{2}$Departamento de Astrof\'{\i}sica Molecular e Infrarroja, Instituto de Estructura de la Materia, CSIC, Serrano 121, E-28006 Madrid, SPAIN}

%
%

%
%
\maketitle
%
\begin{abstract} 

Observational constraints provided by high resolution and high sensitivity observations of external galaxies made in the millimeter and submillimeter range have started to put on a firm ground the study of extragalactic chemistry of molecular gas. In particular, the availability of multi-species and multi-line surveys of nearby galaxies is central to the interpretation of existent and forthcoming millimeter observations of the high redshift universe. Probing the physical and chemical status of molecular gas in starbursts and active galaxies (AGN) requires the use of specific tracers of the relevant energetic phenomena that are known to be at play in these galaxies: large-scale shocks, strong UV fields, cosmic rays and X-rays.  We present below the first results of an ongoing survey, allying the IRAM 30m telescope with the Plateau de Bure interferometer(PdBI), devoted to study the chemistry of molecular gas in a sample of starbursts and AGN of the local universe. These observations highlight the existence of a strong chemical differentiation in the molecular disks of starbursts and AGN.

\end{abstract}

\section{Introduction}

The first millimeter observations of external galaxies used CO as a privileged tracer of molecular gas and were done with single-dish telescopes. These first observations certainly gave essential information on the mass, as well as on the overall distribution and kinematics of molecular gas in galaxies, but they were also plagued by their insufficient spatial resolution and limited coverage of molecular species (essentially CO) (e.g., see Young et al.~1995$^{[1]}$). The pioneering work of Mauersberger, Henkel and collaborators using the IRAM 30m telescope to systematically probe the complex molecular inventory of nearby galaxies (started back in 1989) helped to open our view to molecular species other than CO (e.g., see Mauersberger \& Henkel~1993$^{[2]}$). The advent of highly sensitive mm-arrays has made possible to zoom in on the scales of individual Giant Molecular Clouds (GMCs) or Giant Molecular Associations (GMAs) in nearby galaxies, and provide a complete view of the physical and chemical status of their dense molecular gas reservoirs by going beyond coarse CO mapping.  The spectacular energies injected in the gas reservoirs of starbursts and Active Galactic Nuclei (AGN) can create a particularly harsh environment for the neutral interstellar medium (ISM). This new plethora of multi-species observations can track down the relevant energetic phenomena that are at work all the way along the evolutionary track of starbursts and AGN: large-scale shocks, strong UV fields, cosmic rays and X-rays.

It is worth reminding that the correct interpretation of mm-observations of the high-z universe galaxies, to be made with future mm interferometers (e.g., ALMA), will depend on the availability of multi-wavelength studies made in a number of local universe galaxies used as {\it templates}. In particular, detailed studies of starbursts in the local universe are a prerequisite to interpret how feedback processes driven by star formation and activity may operate at higher redshift galaxies (Springel \& Hernquist~2003$^{[3]}$).

We report below on the results of a combined survey using the IRAM 30m telescope and the Plateau de Bure interferometer (PdBI) devoted to study the chemistry of molecular gas in a sample of starbursts and AGN of the local universe. These observations are adapted to probe the onset of large-scale shocks and the influence of UV fields and X-rays on the chemistry of molecular gas in a sample of nearby prototypical starbursts and AGN such as M~82, NGC~253, NGC~1068 and IC~342. We also discuss the first results of the extension of this survey to a sample of 16 Luminous and Ultraluminous Infrared Galaxies (LIRGs and ULIRGs). 
While the interpretation of these data is still subject to major uncertainties (discussed below), these observations reveal that there is a strong chemical differentiation in the molecular gas reservoirs of starbursts and AGN. 

\begin{figure}[htp!]
   \centering
 \includegraphics[width=14cm]{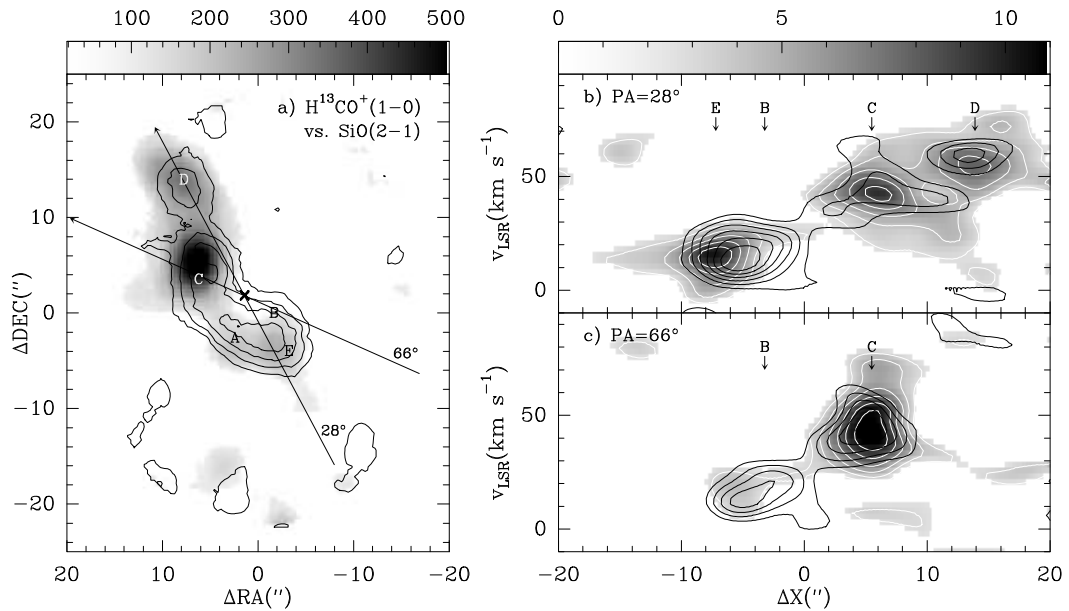}
      \caption{ SiO(2--1) and H$^{13}$CO$^+$(1--0) p--v diagrams 
({\bf b} and {\bf c}, respectively) taken along the 1D--strips highlighted in panel {\bf a} 
where we overlay the SiO(2--1) (grey scale) and H$^{13}$CO$^+$(1--0) (contours) intensity maps of IC~342. The p--v diagrams illustrate the contrast between the different linewidths measured for SiO and H$^{13}$CO$^{+}$ in the northern spiral arm and in the nuclear ring (see discussion of Usero et al.~2006$^{[10]}$).}
       \label{figure_ic342}
   \end{figure}

\section{Chemistry in nearby starburst galaxies and AGN}
\subsection{Molecular shocks in galaxies}

The first SiO($v=0$, $J=2-1$) maps made with the PdBI in the nuclei of the
prototypical starbursts NGC~253 and M~82 have revealed the existence of large-scale
molecular shocks in galaxy disks (Garc\'{\i}a-Burillo et al.~2000$^{[4]}$, 2001$^{[5]}$).
Different scenarios have been brought forward to explain the emission of SiO in our own Galaxy and in the nuclei of external  galaxies. On small $\sim$pc--scales, studies in the galactic disk show that the enhancement of SiO in gas phase can be
produced in the bipolar outflows of young stellar objects (YSOs), due to the sputtering of dust grains by
shocks (Mart\'{\i}n-Pintado et al.~1992$^{[6]}$; Schilke et al.~1997 $^{[7]}$). On larger scales, Mart\'{\i}n-Pintado et al.(1997)$^{[8]}$ detected a SiO $\sim$150~pc circumnuclear disk (CND) in the Galactic Center region. In this CND high fractional abundances of SiO are found in molecular clouds which are not actively forming stars, but where bar models for our Galaxy predict a high likelihood for cloud collisions (H\"uttemeister et al.~1998$^{[9]}$). In M~82, virtually all of the SiO emission traces the disk-halo interface where episodes of mass injection are building up the gaseous halo (Garc\'{\i}a-Burillo et al.~2001$^{[5]}$). Garc\'{\i}a-Burillo et al.(2000)$^{[4]}$ have discussed the role of bar resonances at inducing shocks in  the $\sim$600~pc CND of NGC~253. In the case of our Galaxy but also in NGC~253 (a close to edge-on galaxy) the link between SiO emission and molecular shocks induced by orbit crowding could only be established indirectly via kinematical models due to the high inclination of the two systems.

More recently, the high-resolution images showing the emission of SiO in the inner r$\sim$200~pc disk of IC~342, a barred system with a close to face on orientation, have helped to unambiguously attribute the enhancement of SiO to the onset of large-scale molecular shocks driven by the bar potential of this galaxy (Usero et al.~2006$^{[10]}$).  The
SiO--to--H$^{13}$CO$^+$  intensity ratio measured in IC~342 is seen to increase by an order of magnitude from the inner nuclear ring ($\sim0.3$) to the spiral arm region of the galaxy ($\sim3.3$), i.e., on scales of $\sim$100-200~pc. Most remarkably, the 3mm continuum emission in IC~342, dominated by thermal free-free bremsstrahlung, is mostly anticorrelated with the observed distribution of SiO clouds in the disk. This strongly suggests that large-scale shocks in IC~342 are mostly unrelated with ongoing star formation and that they seem to arise instead in a pre-starburst phase. Moreover, the gas kinematics show significant differences between
SiO and H$^{13}$CO$^+$ over the spiral arm where the linewidths of SiO are a factor of 2 larger than those of
H$^{13}$CO$^+$ (see Fig.~\ref{figure_ic342}). The observed kinematics indicate that molecular shocks arise during cloud-cloud collisions at the stage when kinetic energy has partly dissipated in turbulent motions. The rate of energy dissipated in the shocks is estimated to be comparable to the corresponding rate of energy typically transferred by gravity torques in barred galaxies (Garc\'{\i}a-Burillo et al.~2005$^{[11]}$). The two processes may contribute with comparable weight to drain the energy of the gas on its way to the nucleus. 
While it is true that the inflow of gas in galaxy nuclei is mostly constrained by the angular momentum transfer rather than by the
energy dissipation rate (draining angular momentum is more difficult), this estimate suggests that 
large-scale shocks have indeed a non-negligible influence. These observations have helped to shed light on how density waves
can operate on a clumpy medium such as molecular gas to create shocks and drive gas inflow to the nuclei of galaxies, i.e., 
a good illustration of how chemistry can help to tackle dynamical studies.

Taken together, these results underline the relevant role that large-scale molecular shocks can play at shaping the evolution of gas disks. High-resolution SiO imaging is key to discern the different sources of shock chemistry which are activated at different locations and at different moments in galaxy disks during a starburst event.  More than being a mere tracer of {\it exotic} chemistry, SiO allows to unambiguously probe the regions where dust grains are being destroyed in galaxies due to the action of density waves, star formation and galactic outflows.

\subsection{PDR chemistry in starbursts: the M~82 template}

\begin{figure}[htbp!]
   \centering
.\vspace{-4cm}
 \includegraphics[width=15cm]{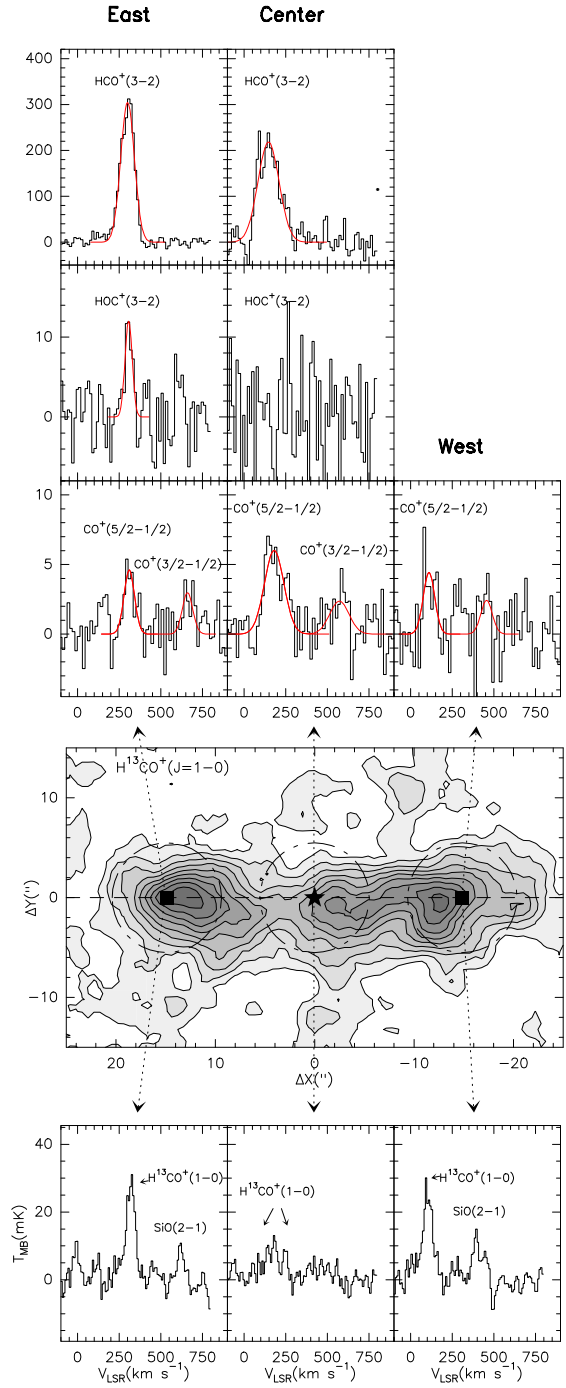}
 \vspace{-2cm}
      \caption{CO$^+$, HOC$^+$ and HCO$^+$ spectra obtained with the IRAM 30m telescope towards 3 positions (named East, West and Center offsets) in the nuclear disk of M~82. The 30m beam patch at 236 GHz is overlaid on the interferometric H$^{13}$CO$^+$ image of Garc\'{\i}a-Burillo et al.(2002)$^{[14]}$ at the three observed positions along the M~82 disk.}
       \label{figure_m82}
   \end{figure}

M~82 is one of the nearest and brightest starburst galaxies which has been studied in virtually all wavebands.
There is mounting evidence that the pervasive UV field produced by the starburst episode has heavily influenced the interstellar medium in M~82 (e.g., Lord et al.~1996$^{[12]}$; Mao et al.~2000$^{[13]}$).

Using the imaging capabilities of the PdBI, Garc\'{\i}a-Burillo et al.(2002)$^{[14]}$ detected widespread emission of the formyl radical (HCO) in the nuclear disk of M\,82. Although observations of this molecule in galactic sources are still too scarce, the emission of HCO is commonly attributed to be enhanced ($\geq$10$^{-10}$) in photo-dissociation regions (PDR). The surprisingly high overall abundance of HCO in M\,82 ($\sim$4$\times$10$^{-10}$) indicates that its nuclear disk has become a giant PDR of $\sim$650\,pc size. Moreover, the highly structured 5$^{\scriptscriptstyle\prime\prime}$ HCO map of M\,82 shows a ring-like distribution, also displayed by other ISM tracers in M\,82. In particular, rings traced by HCO, CO and HII regions appear nested,  with the HCO ring lying in the outer edge of the molecular disk.  The existence of a nested ring pattern, with the highest HCO abundance in the outer ring, suggests that PDR chemistry is {\it propagating} in the disk. Quite puzzling, the spatial correlation between HCO-enriched clouds and HII regions is poor on small scales: offsets of 50--100~pc are found between the peaks of HCO and HII emission. This weak correlation suggests that UV fields coming from the strongest HII regions of M\,82 have photodissociated the bulk of HCO in the envelopes of the closest molecular clouds. Our results suggest that evaporation/destruction of molecular clouds by UV fields is highly efficient in the nuclear disk of the galaxy.  This efficiency shows spatial trends within the M\,82's disk but the reasons 
behind these variations remain to be understood, however.

To have a further insight into the PDR chemistry of M~82, Fuente et al.(2005)$^{[15]}$ mapped the nucleus of this galaxy  in several mm-lines of 
CN, HCN, C$_2$H, c-C$_3$H$_2$, CH$_3$C$_2$H, HC$_3$N  and HOC$^+$ using the 30m telescope. This set of molecular species, including
radicals (CN, C$_2$H), reactive ions (CO$^+$, HOC$^+$) and small hydrocarbons, (c-C$_3$H$_2$) is known to probe the atomic to molecular transition in PDRs. In particular, a [CN]/[HCN] ratio $>$1 has been used as a PDR indicator in galactic regions with very different physical conditions (Fuente et al.~1993$^{[16]}$; Sternberg \& Dalgarno.~1995$^{[17]}$). The 30m data of M~82 show a systematically high [CN]/[HCN] ratio ($\sim$5) in all observed positions across the disk of the galaxy. These results have been interpreted by  Fuente et al.(2005)$^{[15]}$ using the PDR models of Le Bourlot et al.(1993)$^{[18]}$. Adopting G$_0\sim$10$^4$ and a total hydrogen nuclei density $\sim$4$\times$10$^5$~cm$^{-3}$, the model predicts that  [CN]/[HCN] ratios $\sim$5 are only reached in the most heavily UV exposed layers of a PDR (A$_v<$5--6 mag) . Furthermore, Fuente et al.(2005)$^{[15]}$ detected the HOC$^+$(1--0) line with an intensity similar to that of the H$^{13}$CO$^+$(1--0) line. This implies a [HCO$^+$]/[HOC$^+$] ratio of $\sim$ 40. A close examination of the CO$^+$/HCO$^+$/HOC$^+$ chemical network shows that a high ionization degree, X(e$^-$)$>$ 10$^{-5}$, is required to reach [HCO$^+$]/[HOC$^+$]$<$80 (Usero et al.~2004$^{[19]}$). In the frame of the M82 PDR model, this electron abundance is reached at A$_v<$4~mag. This sets a stringent limit to the cloud sizes of the giant PDR of M~82.

More recently, this scenario has received further support after the detection of the reactive ion CO$^+$ in M~82 reported by Fuente et al.(2006)$^{[20]}$ (Fig.~\ref{figure_m82}) . The new 30m data, including the first secure detection of this short-lived ion in 
an external galaxy, corroborate that the molecular gas reservoir in the M~82 disk is heavily affected by the 
UV radiation from the recently formed stars. The [CO$^+$]/[HCO$^+$] ratio is $>$0.04 all across the
M~82 disk. This is one of the largest values of the [CO$^+$]/[HCO$^+$] ratio measured thus far.
While the PDR chemical model adapted by Fuente et al.(2006)$^{[20]}$ to account for the new observations successfully fits the [CO$^+$]/[HCO$^+$] ratios, it falls short of explaining the large CO$^+$ column densities derived for M~82 ($\sim$1--4$\times$10$^{13}$~cm$^{-2}$) by more than one order of magnitude. The failure of PDR models at producing the CO$^+$ column densities derived in prototypical
galactic PDR is a long standing problem, here revealed also on extragalactic scales. The chemistry of reactive ions is very sensitive
to the unknown gas physical conditions in the HI/H$_2$ transition layer. In particular, the gas kinetic temperature profile is poorly
constrained in the high ionization layer of the PDR. A temperature increase in the HI/H$_2$ interface could dramatically rise the production of CO$^+$ and this could partly solve the referred inconsistency between observations and models.

These studies, so far focused on M~82, suggest that UV-rays could be shaping the chemistry of molecular gas of starburst galaxies. The M~82 maps made in these different molecular tracers do not appear as mere scaled versions of each other, a result that underscores the ability of current mm interferometers to efficiently probe chemical differentiation in nearby galaxies. It is worth reminding that extreme starbursts are postulated to be more likely found in the high-redshift universe. The study of local templates such as M~82 is mandatory in order to interpret the future mm-observations to be made in higher redshift galaxies.

\subsection{XDR chemistry in AGN}

AGN can inject vast amounts of energy into their host galaxies through strong radiation fields and rapidly moving jets. It is thus plausible to hypothesize that the feedback influence of activity on the gas reservoir around the central engines of AGN should be disruptive. In particular, molecular gas close to the central engines of active galaxies can be exposed to a strong X-ray irradiation. While the accretion disks of AGN are strong UV emitters, the bulk of the UV flux can be attenuated by neutral gas column densities of only N(H)$\sim$10$^{21}$cm$^{-2}$. Hard X-ray photons (2--10~keV) can penetrate neutral gas column densities out to N(H)$\sim$10$^{23}$-10$^{24}$cm$^{-2}$, however. Therefore, X-ray dominated regions (XDR) could become the dominant sources of emission for molecular gas in the harsh environment of the CND of AGN, as originally argued by Maloney et al.(1996)$^{[21]}$ and Lepp \& Dalgarno~(1996)$^{[22]}$.

\begin{figure}[h]
   \centering
 \includegraphics[width=14cm]{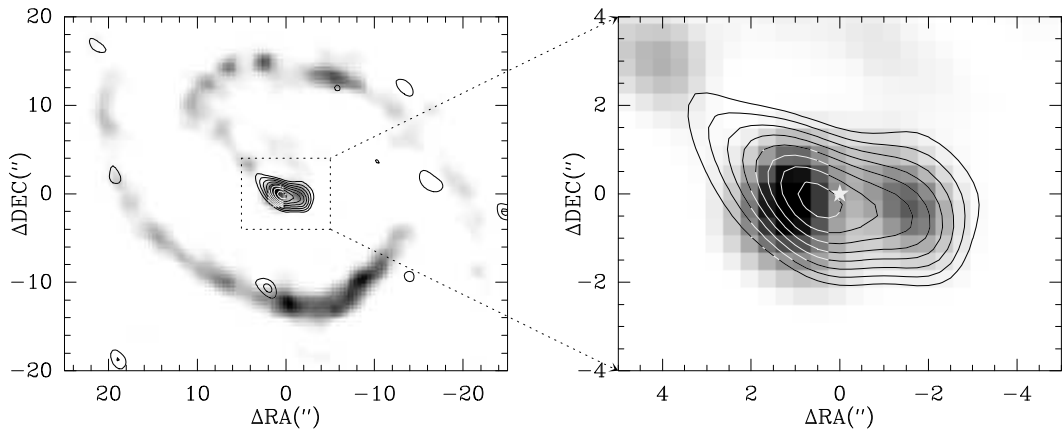}
      \caption{The left panel shows the PdBI SiO(2--1) map of NGC~1068 (line contours; adapted from Garc\'{\i}a-Burillo et al. 2006$^{[27]}$), superposed on the CO(2--1) map (grey scale, adapted from Schinnerer et al.~2000$^{[28]}$). Strong SiO(2--1) thermal emission is associated with the circumnuclear disk around the central engine of the galaxy, while little SiO emission is detected towards the starburst ring. A subset of the SiO map is displayed in the right panel.}
       \label{figure_ngc1068}
\end{figure}

First evidence that the chemistry of molecular gas in the CND of AGN departs from normalcy came from the large HCN/CO luminosity ratio measured in the nucleus of the Seyfert 2 galaxy NGC\,1068 (Tacconi et al.~1994$^{[23]}$). NGC\,1068 has a 
starburst ring of $\sim$2.5-3\,kpc--diameter; significant CO emission arises also from a 200\,pc CND of 
M(H$_2$)$\sim$5$\times$10$^{7}$M$_{\odot}$ (see Fig.~\ref{figure_ngc1068}). The CND, partly resolved into two knots, surrounds the position of the active nucleus. According to the analysis of Sternberg et al.(1994)$^{[24]}$, the high HCN/CO intensity ratio measured by Tacconi et al.(1994)$^{[23]}$ ($\sim$1--10) leads to an abnormally high HCN/CO abundance ratio in the nucleus of NGC~1068: N(HCN)/N(CO)$\sim$a few~10$^{-3}$-10$^{-2}$.

Different scenarios discussed in the literature have tried to find a link between the anomalous HCN chemistry and the presence of an active nucleus in NGC~1068. In particular, the selective depletion of gas-phase oxygen in the dense 
molecular clouds around the AGN would explain the high HCN-to-CO abundance ratio (Sternberg et al.~1994$^{[24]}$;
Shalabiea \&  Greenberg~1996$^{[25]}$). This oxygen depletion scheme predicted a lower-than-normal abundance of all 
oxygen-bearing species. Alternatively, an increased X-ray ionization of molecular clouds near the AGN could enhance the 
abundance of HCN as well as that of other molecular species (e.g., CN and NO), including reactive ions (Maloney et al.~1996$^{[21]}$; Lepp \& Dalgarno~1996$^{[22]}$). Going beyond pure gas-phase chemistry schemes, it has also been argued that X-rays could evaporate small ($\sim$10~\AA) silicate grains, increasing the fraction in gas phase of some refractory elements and subsequently enhancing the abundance of some molecules (e.g., SiO) in X-ray irradiated molecular gas (Voit~1991$^{[26]}$).

To shed some light onto the 'obscuring torus chemistry' of NGC~1068, Usero et al.(2004)$^{[19]}$ conducted observations with the 30m telescope for eight molecular species, purposely chosen to explore the predictions of XDR models for molecular gas. Observations included several lines of SiO, CN, HCO, H$^{13}$CO$^{+}$, H$^{12}$CO$^{+}$, HOC$^{+}$, HCN, CS and CO. 
As shown by Usero et al.(2004)$^{[19]}$ the new estimates of the HCN/HCO$^{+}$ and CN/HCN abundance ratios in the CND of NGC~1068 are  satisfactorily explained by XDR models whereas oxygen-depletion schemes fail to account for them.
In addition, Usero et al.(2004)$^{[19]}$ reported on the detection of significant SiO(3--2) and SiO(2--1) emission towards the central offset of the 30m map of the galaxy, i.e., a position nominally coincident with the CND of NGC~1068. While SiO emission is tentatively detected over the starburst ring in the 30m map, the SiO abundances estimated there are at least 10 times lower than those measured towards the CND. 
The new high-resolution SiO map later obtained with the PdBI (Garc\'{\i}a-Burillo et al. 2006$^{[27]}$) beautifully shows that the bulk of the SiO emission in NGC~1068 comes indeed from the CND with little, if any, contribution coming from the star forming ring (see Fig.~\ref{figure_ngc1068}). The large overall abundance of SiO measured in the CND ($\sim$(5-10)$\times$10$^{-9}$) allowed us to definitely discard the oxygen-depletion scenario. On the other hand, the enhancement of SiO cannot be attributed to the action of shocks driven by star formation on molecular gas as there is hardly any evidence of a recent starburst in the nucleus of NGC~1068. The processing of 10~\AA  dust grains 
by X-rays, as a mechanism to enhance silicon chemistry in gas phase, would explain the large SiO abundances of the CND.  

As a second piece of the puzzle, Usero et al.(2004)$^{[19]}$ also reported on the first extragalactic detection of the reactive ion 
HOC$^+$ in NGC~1068. Most remarkably, the estimated HCO$^+$/HOC$^+$ abundance ratio in 
the nucleus of NGC~1068, $\sim$30--80, is the smallest ever measured in molecular gas. The low  HCO$^+$/HOC$^+$ ratios measured in the CND can be explained if molecular clouds have the high ionization degrees typical of XDR (X(e$^-$)$\sim 
10^{-6}$-$10^{-4}$). An examination of the different formation paths of 
HOC$^{+}$ suggests that reactions involving H$_2$O and/or CO$^{+}$ would be 
the predominant precursors of HOC$^{+}$ in XDR. 

The first global analysis of the combined survey suggests that the bulk of the molecular gas emission in the CND of NGC\,1068 can be best interpreted as coming from a giant XDR created by the central engine.
While shocks related to star formation can not account for the high SiO abundances measured in the CND of NGC\,1068, it remains to be explored whether other mechanisms different to the XDR scenario depicted above can be a valid explanation.
In particular, the onset of large-scale shocks near the inner resonances of the NGC~1068 bar could alternatively explain the enhancement of SiO in the CND, a possible scenario which will deserve further scrutiny based on the new PdBI data.

\section{Extragalactic chemistry in ULIRGs: the way to the high-z universe}

Luminous infrared galaxies (LIRGs, $L_{\rm{IR}} > 10^{11}\,L_{\odot}$) and more particularly their extreme counterparts, ultraluminous infrared galaxies (ULIRGs, $L_{\rm{IR}} > 10^{12}\,L_{\odot}$), have been postulated as the local examples of the higher redshift galaxies which dominate the IR and submm backgrounds (e.g., Blain et al.~1999$^{[29]}$). The nature of the power source of the huge IR emission in these galaxies, an unknown combination of dust enshrouded star formation and obscured accretion onto a black hole (AGN), is still highly debated, however 
(e.g., Sanders et al.~1988$^{[30]}$; Genzel et al.~1998$^{[31]}$; Veilleux et al.~2002$^{[32]}$; Risaliti et al.~2005$^{[33]}$).
LIRGs and ULIRGs are known to possess large amounts of molecular gas (Sanders et al.~1991$^{[34]}$). In addition, Sanders et al.(1988)$^{[30]}$ reported that the infrared--to--CO(1--0) luminosity ratio in ULIRGs is anomalously high compared to that of normal galaxies. This 
remarkable result was interpreted as supporting evidence of the AGN power source scenario for ULIRGs. To shed light on the powering source controversy, Gao \& Solomon~(2004)$^{[35]}$ used HCN(1--0) observations to probe more specifically the dense molecular gas content of a sample of 65 nearby galaxies, including 25 LIRGs and 6 ULIRGs. Their results show that there is an excellent linear correlation between the IR and HCN luminosities over 3 orders of magnitude in $L_{\rm{IR}}$. This correlation, clearly improving that observed between IR and CO(1--0) for $L_{\rm{IR}} > 10^{12}\,L_{\odot}$, was taken as evidence of star formation as being the main power source in ULIRGs. In Gao \& Solomon's view, ULIRGs just happen to host a higher fraction of dense molecular gas compared to normal galaxies and the need of an AGN power source for their $L_{\rm{IR}}$ apparently vanishes. 

The reliability of HCN as a {\it true} tracer of dense molecular gas in LIRGs and ULIRGs has been questioned on several fronts, however. 
First, as argued lines above, X-rays are suspected to significantly enhance HCN abundances in enshrouded AGN (Maloney et al.~1996$^{[21]}$; Kohno et al.~2001$^{[36]}$; Usero et al.~2004$^{[19]}$). Furthermore the excitation of HCN lines in LIRGs and ULIRGs might be affected by IR pumping through a 14\,$\mu$m vibrational transition (Aalto et al.~1995$^{[37]}$).

The possible caveats on the use of {\it only HCN observations} called for the use of alternative tracers of dense gas in LIRGs and ULIRGs. This question is central to disentangling the different power sources of the huge infrared luminosities of these galaxies. This is precisely the purpose of the recent survey conducted by Graci\'a-Carpio et al.(2006)$^{[38]}$ with the IRAM 30m telescope in the 1--0 line of HCO$^+$ of a sample of 16 galaxies including 10 LIRGs and 6 ULIRGs.  The HCO$^+$ and HCN J=1--0 lines have both comparable critical densities and the bulk of their emission is expected to arise from dense molecular gas of n(H$_{2}$) $\geq 10^{4}$\,cm$^{-3}$. To compare the luminosity ratios derived for LIRGs and ULIRGs with those of {\it normal} galaxies, we have compiled a sample of 69 objects including those galaxies for which HCN, HCO$^{+}$ and CO data are available from various sources.

\begin{figure}[h]
   \centering
 \includegraphics[width=14cm]{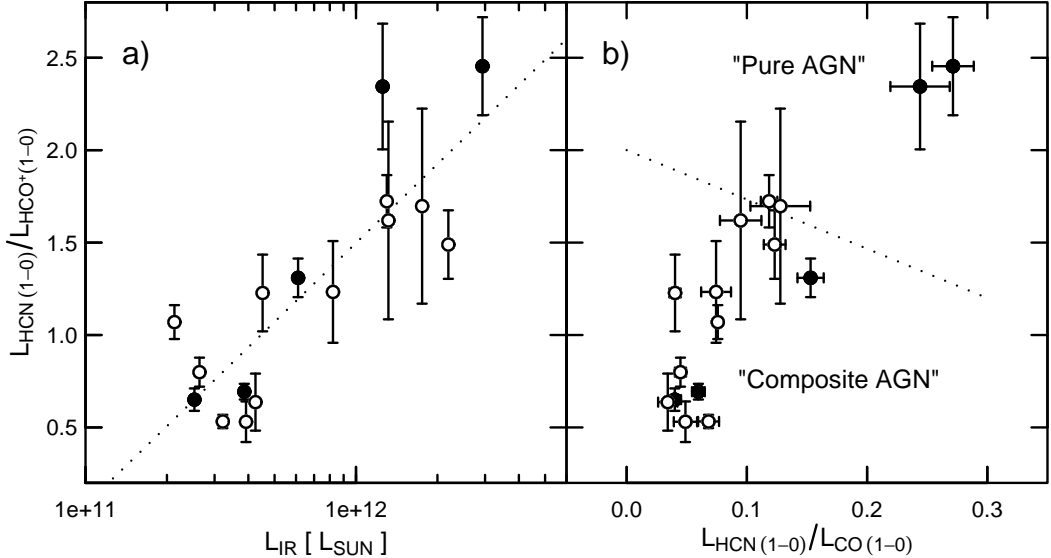}
      \caption{{\bf a)} The correlation between the HCN(1--0)/HCO$^{+}$(1--0) luminosity ratio and $L_{\rm{IR}}$ is displayed for the sample of LIRGs and ULIRGs analyzed by Graci\'a-Carpio et al.(2006)$^{[38]}$. {\bf b)} We show the location of LIRGs and ULIRGs in the 2D diagnostic diagram of Kohno~(2005)$^{[40]}$, distinguishing `pure' from `composite' AGN. Filled symbols represent LIRGs and ULIRGs for which secure identification of embedded AGN are obtained in at least two wavelengths.}
       \label{figure_ulirgs}
\end{figure}

Preliminary results of this HCO$^+$ survey, the first ever conducted in LIRGs and ULIRGs, indicate that the HCN/HCO$^+$ luminosity ratio sharply increases with $L_{\rm{IR}}$ for LIRGs and ULIRGs (Fig.~\ref{figure_ulirgs}). This intriguing and a priori unexpected trend provides indicative evidence that HCN  may not be not a straightforward tracer of dense gas in the most extreme LIRGs.  
The variation of the HCO$^+$(1--0)/CO(1--0) ratio among the sample galaxies shows that the fraction of dense molecular gas of LIRGs and ULIRGs is, on average, a factor of $\sim$2 higher than that of normal galaxies. This is considerably less than the corresponding number derived from HCN (i.e., a factor of $\sim$5). The increase in the dense gas fraction derived from HCO$^+$ would fall short of explaining the observed $L_{\rm{IR}}$ for ULIRGs in the purely star formation scenario of Gao \& Solomon~(2004)$^{[35]}$. Instead, this result suggests that the contribution to $L_{\rm{IR}}$ from an embedded AGN source would amount to $\sim$50\,\% in the most extreme ULIRGs of our sample (Mk~231 and Mk~273). 
Most remarkably, the correlation found between the HCN(1--0)/HCO$^+$(1--0) ratio and  $L_{\rm{IR}}$ for LIRGs and ULIRGs runs in parallel with the long known general tendency of finding more AGN signatures in ULIRGs with increasing $L_{\rm{IR}}$ (Veilleux et al.~1995$^{[39]}$). This seems to be confirmed by the derived location of Mk~231 and Mk~273 in the correlation plot of Fig.~\ref{figure_ulirgs}. In particular, the application to our sample of the diagnostic tool originally designed by Kohno~(2005)$^{[40]}$ to distinguish between `pure' AGN and `composite' starbursts+AGN in nearby Seyferts, reveals that a large number of embedded AGN may be lying in LIRGs and ULIRGs (Fig.~\ref{figure_ulirgs}).

Different mechanisms, either related to the excitation of the HCN(1--0) line, or to the chemical enhancement of the HCN molecule 
can make for HCN(1--0) being over-luminous with respect to HCO$^+$(1--0). The most plausible scenario accounting for the observed trends implies that X-rays shape the chemistry of molecular gas at $L_{\rm{IR}} > 10^{12}\,L_{\odot}$. The future extension of this work to galaxies with higher $L_{\rm{IR}}$ (HyLIRGs and high-$z$ submm sources) will help to shed light on the relative contribution of star formation and AGN to the huge infrared luminosity of high-$z$ objects.  More recently, HCN observations of an ultraluminous quasar at $z = 3.9$ have nicely confirmed that the contribution from an embedded AGN to $L_{\rm{IR}}$ in high-$z$ galaxies could be dominant (Wagg et al.~2005$^{[41]}$). This underscores the ability of multi-species mm-surveys in conjunction with chemical modelling of observations to address the controversial question of what lurks inside 
ULIRGs and high-$z$ submillimeter sources.

\section{Concluding remarks}

The significant progress made in mm radioastronomy during the last decades has made possible 
to start putting on a firm ground the study of extragalactic chemistry.
The interpretation of this plethora of new multi-line/multi-species observations of external galaxies is certainly subject to major uncertainties, however. In particular, the unknown beam filling factor of the effective emitting regions of extragalactic sources as observed in different molecular lines, makes the translation of the observed intensity ratios into {\it useful} numbers (e.g., physical conditions:--density, temperature--and chemical abundances) somewhat risky. 
In some cases, the problem is degenerate, and a single set of intensity ratios derived from low resolution maps can be simultaneously explained by an adequate combination of different excitation conditions and/or different chemical abundances of the observed molecular species. In this case, the comparison with models is a tricky business. This degeneracy starts to be resolved when the spatial resolution of observations is significantly boosted for a large number of molecular lines of different molecular species. In the case of starbursts and AGNs, interferometer maps, zooming in on the scales of individual GMCs or GMAs, clearly reveal that there is a strong large-scale chemical differentiation in the molecular gas reservoirs of these galaxies.

Different processes can shape the evolution of the molecular gas reservoirs of galaxies along the typical evolutionary track of a starburst: large-scale shocks, strong UV-fields, cosmic-rays and also X-rays (the latter are mostly relevant when an activity episode is switched on). These processes are expected to be at work at different stages (from the pre-starburst phase to the phase of starburst remnant), but also, at different locations in the disks, in the halos and in the disk-halo interfaces of galaxies.
The use of specific tracers of these relevant energetic phenomena can help to track down the evolution of galaxies.

While extragalactic chemistry is a useful tool to understand galaxy evolution, these new extragalactic data provide  
major constraints but also pose in return major problems to chemical modelers. Below, we itemize a short list of open questions, naturally biased by the new results presented above:

\begin{itemize}

\item 

The SiO molecule has revealed to be a privileged tracer of shocks in galaxies. Surprisingly (?), these shocks happen to be
rather unrelated with on-going massive star formation in galaxies, but more indicative of a pre-starburst phase (e.g., density waves in NGC~253 and IC~342) or of an evolved phase of the starburst (e.g., the disk-halo interface in M~82). Unfortunately, these observations can not provide accurate estimates of the mass and typical velocities of the molecular gas actually involved in these shocks, both being most relevant numbers. One of the underlying reasons is that current models of SiO production in interstellar shocks do not easily distinguish between the case of Si-bearing material in the core and in the mantle, and thus, the energetics of the shock process is not well constrained. 
Nor is there a clear picture of what may be the identity of sputtering products in either case. The use of alternative tracers of molecular shocks different to SiO (SO?, CH$_3$OH?, HNCO?, SO$_2$?,...) could help to disentangle the issue. However, some of these tracers are suspected to be also affected by photo-sputtering and/or evaporation, and thus may not provide clear cut cases. Complementary observations probing the size, temperature and poorly known composition of dust grains in galaxies (doable with SPITZER) will also shed light on this problem.

\item 

The observations of M~82 have revealed a giant PDR of 650\,pc--size. Most remarkably the HCO map of M~82 shows indications that
PDR chemistry is propagating outwards in the disk of the galaxy. However, the large abundances measured for HCO in M~82 (X(HCO)$\sim$a few 10$^{-10}$) cannot be accounted for by current schemes of HCO production in PDRs. These models, invoking the photo-sputtering of H$_2$CO from dust grains and the ulterior photo-dissociation of this molecule in gas-phase giving HCO as a yield, typically foresee that X(HCO) should be $\sim$10$^{-11}$, i.e., an order of magnitude less than the value estimated in M~82 on scales of 650pc!. The HCO results in M\,82 lead us to
underscore the need of including dust grain chemistry in PDR models. HCO is not the only molecule to be apparently 'over-produced' on large scales in the giant PDR of M~82. The large abundances estimated in M~82 for molecular species and reactive ions such as CO$^+$, HOC$^+$ and some small hydrocarbons (e.g., c-C$_3$H$_2$) are a real challenge for the validity of current PDR models. 
In the case of CO$^+$ or HOC$^+$, the predictions of PDR models on their abundances could improve if we had a better handle on the likely steep temperature profile of the HI/H$_2$ interface. The large abundance of small hydrocarbons in PDRs have been tentatively explained as a result of the destruction of PAHs, a scenario which is still under discussion.    

\item

XDR could become the dominant sources of emission for molecular gas in the circumnuclear playground of AGNs.  
Diagnostic tools supplied by modelers should be confronted to mm-line observations in order to possibly distinguish between the XDR and PDR scenarios. This is especially relevant in the case of deeply embedded AGNs whose identification cannot be easily made at higher frequencies due to the prevalent dust obscuration. Of particular note, obscured AGNs and/or embedded starbursts are suspected to be more frequently found in the early universe. The future development of new XDR models should allow to confront observations in the mm-range of high-redshift galaxies with the expected output of models for some critical molecular species.
With the exception of the data discussed in this paper, knowledge of the dense gas content of ULIRGs (and of high redshift galaxies) had thus far been limited to that emerging from only--HCN--line studies. While HCN is expected to be overly produced in XDR (and also in PDR), alternative tracers should be observed in order to distinguish between the PDR and XDR cases.

\end{itemize}




\end{document}